\documentstyle[12pt]{article}
\begin{document}
\title{Chiral Anomaly and $\gamma 3\pi^*$}
\author{Barry R. Holstein\\
Department of Physics and Astronomy\\
University of Massachusetts\\
Amherst, MA  01003}
\maketitle
\begin{abstract}
Measurement of the $\gamma 3\pi$ process has revealed a possible conflict with
what should be a solid prediction generated by the chiral anomaly.  We show
that inclusion of appropriate energy-momentum dependence in the matrix element
reduces the discrepancy.
\end{abstract}
\vfill
{}$^*$ Research supported in part by the National Science Foundation.
\newpage

\section{Introduction}
The chiral anomaly is a well-known and fascinating aspect of QCD.  First
identified in the context of the ``triangle diagram" contribution to
$\pi^0\rightarrow 2\gamma$,\cite{1} it has been shown to have much more general
consequences which can be characterized in terms of an effective
Lagrangian\footnote{We include here only the component relevant to
electromagnetic
interactions.}\cite{2}
\begin{eqnarray}
{\cal L}_{WZW}={N_c\over 48\pi^2}\epsilon^{\mu\nu\alpha\beta}[
eA_\mu\mbox{Tr}(Q(R_\nu R_\alpha R_\beta+L_\nu L_\alpha L_\beta))\nonumber\\
-ie^2F_{\mu\nu}A_\alpha\mbox{Tr}(Q^2(L_\beta+R_\beta)+{1\over 2}
(QU^\dagger QUR_\beta+QUQU^\dagger L_\beta))]
\end{eqnarray}
where $U=\exp(i\sum \lambda_i\phi_i/F_\pi)$ is the usual nonlinear
matrix describing the pseudoscalar Goldstone fields,
$R_\mu\equiv(\partial_\mu U^\dagger)U,\qquad L_\mu\equiv U(\partial
U^\dagger)$ are right, left-handed currents respectively
and $Q={e\over 3}(2,-1,-1)_{\rm diag}$ is the quark charge matrix.
One immediately identifies the theoretical prediction for
$\pi^0\rightarrow\gamma\gamma$ which arises from the second line of Eq. 1
\begin{eqnarray}
Amp_{\pi\rightarrow\gamma\gamma}=-iA_{\gamma\gamma}
\epsilon^{\mu\nu\alpha\beta}\epsilon^*_\mu k_\nu{\epsilon'}^{*}_\alpha
{k'}_\beta
\nonumber\\
\quad\mbox{with}\qquad A_{\gamma\gamma}={\alpha N_c\over 3\pi F_\pi}
\stackrel{N_c=3}\longrightarrow 0.025 \mbox{GeV}^{-1}
\end{eqnarray}
which is in excellent agreement with the experimental value\cite{3}
\begin{equation}
A_{\gamma\gamma}=0.025\pm 0.001 \mbox{GeV}^{-1}
\end{equation}

In a corresponding fashion one can read off from Eq. 1 the prediction for
the $\gamma\pi\pi\pi$ vertex
\begin{eqnarray}
Amp_{\gamma\pi^+\pi^-\pi^0}=-iA_{3\pi}(0)\epsilon^{\mu\nu\alpha\beta}\epsilon_\mu p_{1\nu}p_{2\alpha}
p_{0\beta}\nonumber\\
\quad{\rm with}\quad A_{3\pi}(0)={eN_c\over
12\pi^2F_\pi^3}\stackrel{N_c=3}\longrightarrow 9.7 \mbox{GeV}^{-3}
\end{eqnarray}
In this case, agreement with the value quoted experimentally\cite{4}
\begin{equation}
A_{3\pi}^{\rm exp}=12.9\pm 0.9\pm 0.5\mbox{GeV}^{-3}
\end{equation}
is not particularly convincing and could even be said to favor the value
$N_c=4$!  However, since such a violation would have severe consequences
about the very foundations of QCD it warrants a more careful look, which is
the purpose of the present note.  Since the prediction of the anomaly strictly
speaking hold only at zero four-momentum, while the experimental data is
obtained over a range of energies above threshold, it is essential to
understand the energy dependence of the $\gamma3\pi$ amplitude generated by
${\cal O}(p^6)$ and higher contributions, and this is done in section II.
Then in
section III we use these results to confront existing experimental information
and comment on implications for future experiments such as that
approved at CEBAF.\cite{22}

\section{Finite Energy Corrections}

The issue of finite energy correction to predictions of the anomaly has been
addressed by a number of authors and is now reasonably well understood.  The
first such consideration was that of Terent'ev who, on phenomenological
grounds, suggested the form\cite{5}
\begin{equation}
A_{3\pi}(s,t,u)=A_{3\pi}(0)[1+C_\rho e^{i\delta}({s\over m_\rho^2-s}+{t\over
m_\rho^2-t}+{u\over m_\rho^2-u})]
\end{equation}
where $s=(p_1+p_2)^2, t=(p_1+p_0)^2, u=(p_2+p_0)^2$, $\delta$ is an
phenomenological phase factor, and
\begin{equation}
C_\rho={2g_{\rho\pi\pi}g_{\pi\rho\gamma}\over m_\rho^3A_{3\pi}(0)}=0.478
\end{equation}
represents the pure vector dominance contribution.
The next step was taken by Rudaz who, noting that the amplitude
for $\pi^0\rightarrow\gamma\gamma$ could be generated entirely via the vector
dominance diagram $\pi^0\rightarrow \omega\rho\rightarrow\gamma\gamma$, {\it
cf.} Figure 1a, proposed the same for the $\gamma3\pi$ process, {\it cf.}
Figure 1b, yielding\cite{6}
\begin{equation}
A_{3\pi}(s,t,u)={1\over 3}A_{3\pi}(0)[{m_\rho^2\over m_\rho^2-s}+{m_\rho^2\over
m_\rho^2-t}+{m_\rho^2\over m_\rho^2-u}]
\end{equation}
\begin{figure}
\vspace{2.5in}
\caption{Vector dominance contributions to the reactions $\pi^0\rightarrow
\gamma\gamma$ (a) and $\gamma\rightarrow 3\pi$ (b).}
\end{figure}

However, it was soon realized that this expression conflicted both with
the KSRF relation\cite{10} as well as with the anomalous
Ward identities of Aviv and Zee\cite{7} and that the correct form was\cite{8}
\begin{equation}
A_{3\pi}(s,t,u)=-{1\over 2}A_{3\pi}(0)[1-({m_\rho^2\over
m_\rho^2-s}+{m_\rho^2\over m_\rho^2-t}
+{m_\rho^2\over m_\rho^2-u})]
\end{equation}
which contains both a vector dominance piece {\it and} a contact term.

In recent years, the problem has also been addressed via a one loop expansion
in chiral perturbation theory, yielding the form, correct to ${\cal O}(p^6)$
in the derivative expansion\cite{9}\footnote{Here we use the mass shell
condition
$s+t+u=3m_\pi^2$ and determine the coefficient of the term linear in
s,t,u (a free parameter in strict chiral perturbation theory) by demanding
agreement with expansion of the vector dominance form Eq. 9.}
\begin{equation}
A_{3\pi}(s,t,u)=A_{3\pi}(0)[1+{3m_\pi^2\over 2m_\rho^2}+{m_\pi^2\over
24\pi^2 F_\pi^2}({3\over 4}\ln{m_\rho^2\over m_\pi^2}+F(s)+F(t)+F(u))]
\end{equation}
where
\begin{equation}
F(s)=\left\{
\begin{array}{ll}
(1-{s\over 4m_\pi^2})\sqrt{s-4m_\pi^2\over s}\ln{1+\sqrt{s-4m_\pi^2\over
s}\over -1+\sqrt{s-4m_\pi^2\over s}}-2 & s>4m_\pi^2 \\
2(1-{s\over 4m_\pi^2})\sqrt{4m_\pi^2-s\over s}\tan ^{-1}\sqrt{s\over
4m_\pi^2-s}-2& s \leq 4m_\pi^2
\end{array}\right.
\end{equation}

The vector dominance
form---Eq. 9---may be made consistent with its chiral
counterpart---Eq. 10---provided we include the effects of final state
p-wave pi-pi scattering.  We begin by noting that the N/D form
\begin{equation}
t_1(s)=t_1^{CA}(s)/D_1(s),
\end{equation}
with
\begin{equation}
t_1^{CA}(s)={s-4m_\pi^2\over 96\pi F_\pi^2}
\end{equation}
being the familiar p-wave Weinberg or current algebra prediction\cite{26} and
\begin{equation}
D_1(s)=1-{s\over m_\rho^2}-{s\over 96\pi^2F_\pi^2}\ln{m_\rho^2\over m_\pi^2}-
{m_\pi^2\over 24\pi^2F_\pi^2}F(s)
\end{equation}
providing an analytic approximation to the Omnes function,\footnote{One could
also use the experimental p-wave phase shifts and the definition
\begin{equation}
D_1(s)=\exp \left(-{s\over \pi}\int_{4m_\pi^2}^\infty {ds'\delta_1(s')\over
s'(s'-s-i\epsilon)}\right)
\end{equation}
but the result is similar.}\cite{27}
provides a rather successful representation for the $\ell=1$ pi-pi scattering
amplitude\cite{28}
\begin{equation}
t_1(s)=\sqrt{s\over s-4m_\pi^2}e^{i\delta_1(s)}\sin\delta_1(s).
\end{equation}
Likewise, a reasonable approximation to the electromagnetic form factor of
the charged pion is\cite{29}
\begin{equation}
G_\pi (s)=1/D_1(s)\approx {m_\rho^2\over m_\rho^2-s-im_\rho\Gamma_\rho(s)}
\end{equation}
where
\begin{equation}
\Gamma_\rho(s) = \theta({s\over 4m_\pi^2}-1)
{g_{\rho\pi\pi}^2s\over
48\pi m_\rho}\left(1-{4m_\pi^2\over s}\right)^{3\over 2}
\end{equation}
is an energy dependent quantity which reduces to the rho width when
$s=m_\rho^2$.  Here we have noted that
\begin{equation}
{m_\pi^2\over 24\pi^2F_\pi^2}\mbox{Im}F(s)={1\over m_\rho} \Gamma_\rho(s)
\end{equation}
and have utilized the KSRF relation
$g_{\rho\pi\pi}^2=m_\rho^2/2F_\pi^2$.\cite{10}
We observe that Eqs. 9 and 10 can be made to agree to low order in $s,t,u$
provided we use the form
\begin{eqnarray}
A_{3\pi}(s,t,u)&=&-{1\over 2}A_{3\pi}(0)[1-({m_\rho^2\over
m_\rho^2-s}+{m_\rho^2\over m_\rho^2-t}+{m_\rho^2\over m_\rho^2-u})]\nonumber\\
&\times&\left({1-{s\over m_\rho^2}\over D_1(s)}\right)
\left({1-{t\over m_\rho^2}\over D_1(t)}\right)\left({1-{u\over m_\rho^2}\over
D_1(u)}\right)
\end{eqnarray}
which is suggested by the feature that rescattering occurs in each of the
three pi-pi channels simultaneously.  It should also be noted that
Eq. 20 satisfies the requirements of the Fermi-Watson theorem ({\it i.e.}
unitarity) for the
process $\gamma\pi\rightarrow\pi\pi$ and provides
the preferred form to use in future analysis.

\section{Comparison with Experiment}

As mentioned in the introduction, it is often asserted that the experimental
and theoretical values for $A_{3\pi}(0)$ are in significant disagreement.
However, a more careful look at the paper of Antipov et al.\cite{4} reveals
that this is
not the case.  In fact, the experimental value quoted in Eq. 5 obtains only
under the assumption that the matrix element $A_{3\pi}(s,t,u)$ is {\it
independent} of momentum.  On the other hand, averaging the from given by
Terent'ev over the experimental spectrum yields (in units of
GeV$^{-3}$)\cite{4}
\begin{equation}
A_{3\pi}^2(0)+1.9\cos\delta A_{3\pi}(0)+1= 166 \pm 23\pm 13
\end{equation}
Since the spectral shape given by Terent'ev---Eq. 6---is basically in agreement
with the
form given by anomaly considerations--- Eq. 9---provided $\cos\delta =1$, and
since the
experiment of Antipov et al. was primarily at low values of the energy where
unitarity corrections given by Eq. 19 are small we find the
solution
\begin{equation}
A_{3\pi}(0)=11.9\pm 0.9\pm 0.5 \mbox{GeV}^{-3}
\end{equation}
Thus the disagreement with the number required by the chiral anomaly is at the
1.6$\sigma$ level rather than the 2.3$\sigma$ level generally quoted.
Nevertheless, the experimental value is still on the high side and should
certainly be subjected to additional experimental scrutiny,
as will take place in the approved CLAS experiment
at CEBAF.\cite{22}  When such data are analyzed they should use forms
such as Eq. 20
which both satisfy chiral and unitarity restrictions as well as the
phenomenological
requirements of vector dominance.  That use of such a form can make a
significant difference
can be seen in Table 1, where we compare the modifications of the
lowest order anomaly prediction as generated by Eqs. 10,9,20.\footnote{The
top line of each row is equivalent
to the results quoted previously by Bijnens, Bramon and Cornet, ref 9.}
\begin{table}
\caption{Spectral modifications to the process $\gamma\pi\rightarrow\pi\pi$
generated via Eqs. 10,9,20 respectively.  All values of s,t are in units of
$m_\pi^2$ and the numbers quoted in the table represent percentage deviations
from the anomaly prediction.}
\begin{center}
\begin{tabular}{c|c|c|c|c|c}

s,|t| & 0.5 & 1.0 & 2.0 & 3.0& 4.0 \\
\hline
4 &    6.0 & 6.0 & 6.0 & 6.3 & 6.6 \\
  & 5.9 & 5.9 & 5.9 & 6.0 & 7.2 \\
  & 6.9 & 6.9 & 6.9 & 8.5 & 10 \\
5 & 6.6 & 6.6 & 6.6 & 6.7 & 6.8 \\
  & 6.0 & 6.0 & 6.0 & 7.2 & 7.3 \\
  & 8.5 & 8.5 & 8.5 & 10 & 11 \\
10& 8.7 & 8.7 & 8.7 & 8.6 & 8.5 \\
  & 14 & 14 & 14 & 14 & 14 \\
  & 21 & 21 & 21 & 21 & 21 \\
15& 11 & 11 & 11 & 11 & 11 \\
  & 32 & 31 & 30 & 30 & 30 \\
  & 45 & 45 & 44 & 44 & 44 \\
20& 14 & 14 & 14 & 14 & 14 \\
  & 70 & 70 & 70 & 68 & 68 \\
  & 96 & 95 & 95 & 93 & 93 \\
\hline
\end{tabular}
\end{center}
\end{table}
In the region $4m_\pi^2<s<13m_\pi^2; 0.5m_\pi^2<|t|<3.5m_\pi^2$ explored by
the Antipov et al experiment the differences between the various forms are
moderate, but in CEBAF proposal much larger values of energy and momentum
transfer are involved---$4m_\pi^2<s,|t|<50m_\pi^2$ and the use of a properly
unitarized form for the decay amplitude is essential in order to extract the
value of the anomaly.

\end{document}